\documentclass
[%
reprint,
 showpacs,preprintnumbers,
 amsmath,amssymb,
 aps,
 pre,
 floatfix,
 longbibliography
]{revtex4-1}

\usepackage{booktabs}
\usepackage{array}
\newcolumntype{L}[1]{>{\raggedleft\arraybackslash}p{#1}}
\AtBeginDocument{
\heavyrulewidth=.08em
\lightrulewidth=.05em
\cmidrulewidth=.03em
\belowrulesep=.65ex
\belowbottomsep=0pt
\aboverulesep=.4ex
\abovetopsep=0pt
\cmidrulesep=\doublerulesep
\cmidrulekern=.5em
\defaultaddspace=.5em
}

\usepackage{hyperref}
\usepackage{color}
\usepackage{xcolor}
\usepackage{graphicx}% Include figure files
\usepackage{dcolumn}% Align table columns on decimal point
\usepackage{bm}% bold math
\usepackage{cancel}
\usepackage[utf8]{inputenc}

\begin{document}

\title{Out-of-equilibrium Monte Carlo simulations of a classical gas with Bose-Einstein statistics}

\author{M. Di Pietro Mart\'inez}
\author{M. Giuliano}
\author{M. Hoyuelos}
\email{hoyuelos@mdp.edu.ar}

\affiliation{Instituto de Investigaciones F\'isicas de Mar del Plata, CONICET, Facultad de Ciencias Exactas y Naturales, Universidad Nacional de Mar del Plata, Funes 3350, 7600 Mar del Plata, Argentina}

\date{\today}% It is always \today, today,
             % but any date may be explicitly specified

\begin{abstract}
Algorithms to determine transition probabilities in Monte Carlo simulations are tested using a system of classical particles with effective interactions which reproduce Bose-Einstein statistics. The system is appropriate for testing different Monte Carlo simulation methods in out-of-equilibrium situations since non equivalent results are produced. We compare mobility numerical results obtained with transition probabilities derived from Glauber and Metropolis algorithms. Then, we compare these with a recent method, the interpolation algorithm, appropriate for non-equilibrium systems in homogeneous substrata and without phase transitions. The results of mobility obtained from the interpolation algorithm are qualitatively verified with molecular dynamics simulations for low concentrations.
\end{abstract}

%\pacs{05.40.-a, 05.60.-k, 66.10.Cg, 05.10.Gg}

\maketitle

%##################################################################################
\section{Introduction}
If transition probabilities between different states of a system are known, then the kinetic Monte Carlo algorithm can be used to numerically reproduce a correct description of the transient or non-equilibrium behavior \cite{bortz,fichthorn,voter,jansen}. Nevertheless, it is not uncommon that the information available is restricted to the state's energy, and transition probabilities have to be estimated using, for example, Glauber or Metropolis algorithms. Such algorithms guarantee a correct description of the equilibrium state, but not of the out-of-equilibrium transient. Convergence towards equilibrium is ensured by imposing the detailed balance condition on the transition probabilities, see for example~\cite{binder,binder2}.

Let us consider a system of particles, at temperature $T$, divided into cells; transition probabilities describe jumps of one particle between neighboring cells. In Refs.~\cite{suarez,mdp,dipietro2}, it has been shown that the detailed balance condition can be used to derive a class of transition probabilities characterized by an interpolation parameter $\theta$. If $\theta=-1$, the transition probability depends on the one-particle potential in the origin cell; if $\theta=1$, it depends on the potential in the target cell; and if $\theta=0$, it depends on the potential energy change (a frequent choice in Monte Carlo simulations). 
The interpolation algorithm consists in the implementation of the kinetic Monte Carlo method using transition probabilities that are obtained from the potential and the interpolation parameter \cite{suarez,mdp,dipietro2}. The purpose of this paper is to verify that this algorithm correctly describes an out-of-equilibrium system.
To do this we select a specific example.
The chosen system is a gas of classical particles with an effective attractive interaction which reproduces Bose-Einstein statistics. The main reason for this choice is that this system exhibits clear differences among the results generated by the mentioned numerical algorithms, in contrast, as shown below, with purely repulsive interaction of hard core.

So, we consider the following stationary non-equilibrium state. A constant force $F$ is applied along direction $x$; the external potential is $U_i = -F x_i$, where $x_i$ is the position of cell $i$. The system has periodic boundary conditions along the $x$ axis and, after some time, a stationary current is established. Particles have a mean velocity proportional to $F$ and the proportionality constant is the mobility $B$. We consider $B$ (for small values of $F$) as the parameter which characterizes the non-equilibrium state, and analyze its dependence on the density for the different algorithms.

This article is organized as follows. In Sec.\ \ref{sec:alg}, the three algorithms used for transition probabilities: Glauber, Metropolis and interpolation, are described. In Sec.\ \ref{sec:hs}, the hard-core interaction is briefly analyzed. In Sec.\ \ref{sec:basic}, the basic formulae for the Bose-Einstein gas are derived. %, specifically the expression for the configuration energy and the connection between the lattice gas description and the kinetic description.
In Sec.\ \ref{sec:gm}, we present numerical and analytical results of the mobility using Glauber, Metropolis and interpolation algorithms for the Bose-Einstein gas. The Monte Carlo simulations are performed in a one-dimensional lattice gas with periodic boundary conditions.
% it corresponds to a string of cells with periodic boundary conditions.
%In Sec.\ \ref{sec:inter}, we present the results of kinetic Monte Carlo simulations using the interpolation algorithm for the transition probabilities.
For completeness and as a verification, in Sec.\ \ref{sec:md}, we compare the previous results with the ones obtained through Molecular Dynamics. This corresponds to a more detailed kinetic description, where the position and velocity of each particle inside the cells are considered~\cite{supmat}
% In a more detailed kinetic description, we consider position and velocity of each particle inside cells. This is done in Sec.\ \ref{sec:md}, where molecular dynamics results of the mobility are compared with the previous results.
Finally, conclusions are presented in Sec.\ \ref{sec:conclusions}.

\section{Algorithms for transition probabilities}
\label{sec:alg}

According to the Glauber algorithm \cite{glauber}, in a Monte Carlo step the transition from state $i$ to state $j$, with energies $E_i$ and $E_j$ respectively, has a probability
\begin{equation}\label{eq:glauber}
p_{i,j}^G = \frac{1}{1 + e^{\beta(E_j-E_i)}},
\end{equation}
where $\beta=(k_B T)^{-1}$.

In the Metropolis or Metropolis-Hastings algorithm \cite{hastings}, the probability is
\begin{equation}\label{eq:metro}
p_{i,j}^M = \min (1,e^{-\beta(E_j-E_i)}).
\end{equation}

Note that Metropolis algorithm is faster: if $E_i=E_j$, we have $p_{i,j}^G = 1/2$ and $p_{i,j}^M = 1$. We have to take this difference into account to have the same time scale in both cases. If $P$ is the number of jump attempts per unit time, then the transition probabilities are
\begin{align}
W_{i,j}^G &= 2\, P\, p_{i,j}^G \\
W_{i,j}^M &= P\, p_{i,j}^M
\end{align}
where a factor 2 is included in $W_{i,j}^G$ to compensate the speed difference between algorithms.

The energy of a given configuration $\{n_i \}$ is
\begin{equation}\label{eq:energy}
E= \sum_i (\phi_{n_i}+n_i U_i),
\end{equation}
where $\phi_{n_i}$ represents the interaction energy (or configuration energy) among the $n_i$ particles in the cell, in local equilibrium, and $U_i$ is an external potential. Only processes where one particle jumps to a neighboring cell are allowed.

A different perspective is adopted for the interpolation algorithm. Instead of using the state energy, the mean field potential $V_i$ for one particle, produced by all the other particles in the cell, is considered. We call $\theta_i$ the interpolation parameter in cell $i$. The transition probability from cell $i$ to cell $i+1$ is given by
\begin{equation}
W_{i,i+1} = P e^{-\beta\left[ \theta_i V_i + \theta_{i+1} V_{i+1} + \Delta V + \Delta U \right]/2},
\label{eq:transpr}
\end{equation}
where $\Delta U=U_{i+1}-U_i$ and $\Delta V=V_{i+1}-V_i$. See Ref.\ \cite{vattulainen} for a related approach in which the transition probability depends on the sum of energies in origin and target sites. Note that $\theta_i$ and $V_i$ can be functions of the number of particles, $n_i$, in a cell $i$.

A relationship between $V_i$ and $\phi_{n_i}$ has to be established. This can be done using the equilibrium distribution of particles $\bar{n}_i$. From the transition probabilities \eqref{eq:transpr}, a Fokker-Planck equation can be derived, whose equilibrium solution is 
\begin{equation}\label{eq:equil}
\bar{n}_i = e^{-\beta (V_i + U_i - \mu)},
\end{equation}
where $\mu$ is the chemical potential, see Ref.\ \cite{suarez}. On the other hand, $\bar{n}_i$ can also be obtained from the grand partition function for cell $i$ with energy $\phi_{n_i}+n_i U_i$. Combining both results, it can be shown \cite{suarez} that
\begin{equation}
e^{-\beta V_i} = \langle e^{-\beta (\phi_{n_i+1}-\phi_{n_i})}\rangle,
\label{eq:widom}
\end{equation}
where $V_i$ is evaluated at the average number $\bar{n}_i$ ($V_i$ is a continuous function of $\bar{n}_i$ but, when implementing the algorithm for specific realizations, $V_i$ has to be evaluated only at integer values of $n_i$).

The interpolation parameter $\theta_i$ can be simplified in the definition of transition probabilities \eqref{eq:transpr} using the following relationship that holds in the absence of a phase transition:
\begin{equation}\label{eq:etheta}
e^{-\beta \theta_i V_i} = \frac{1}{1 + \beta n_i V'_i},
\end{equation}
where $V'_i$ is the the derivative of $V_i$ with respect to the number of particles. The demonstration of \eqref{eq:etheta} can be found in Ref.\ \cite{dipietro3}; it is obtained from the fact that the energy difference between final and initial states is equal to the difference between the energy barriers (the exponents in $W_{i,i+1}$) of the forward and backward processes. The transition probability form cell $i$ to $i+1$ becomes
\begin{equation}\label{eq:transpr2}
W_{i,i+1} = P \frac{e^{-\beta(\Delta V + \Delta U)/2}}{(1 + \beta n_i V'_i)^{1/2}(1 + \beta n_{i+1} V'_{i+1})^{1/2}}.
\end{equation}
In summary, transition probabilities in the interpolation algorithm are not obtained directly from the state energy, but are functions of the potential $V_i$ that is obtained from \eqref{eq:widom}. The method is more involved than Glauber or Metropolis, and it only applies to system of particles with local interactions, where the energy can be written as in \eqref{eq:energy}. The advantage is that  it holds out of equilibrium.

Here, models where $\phi_{n_i}$ is a simple function of $n_i$ are considered. Long range interactions, such as the Coulomb potential, can not be represented by these methods, where the energy of one cell, given by $\phi_{n_i} + n_i U_i$, does not depend on the number of particles in nearby cells. Interaction between particles in different cells is neglected. For a specific (short range) interaction potential, the configuration energy $\phi_{n_i}$ in general has to be obtained numerically. If the interaction potential is  $\mathcal{U}_i(\mathbf{q}_1,\dots,\mathbf{q}_{n_i})$, with $\mathbf{q}_j$ the position of particles in cell $i$, the configuration energy $\phi_{n_i}$ is given by 
\begin{equation}\label{eq:confe}
e^{-\beta \phi_{n_i}} = \langle e^{-\beta \mathcal{U}_i(\mathbf{q}_1,\dots,\mathbf{q}_{n_i})} \rangle^0
\end{equation}
where average $\langle \  \rangle^0$ is computed with the probability distribution of non-interacting particles (that is, particles randomly distributed).

\section{Hard-core interaction}
\label{sec:hs}

The main purpose of this paper is to check the different algorithms with an effective Bose-Einstein interaction. But before that, we wish to briefly analyze other simple potential: hard core. These potentials represent two extreme situations: purely attractive or purely repulsive interactions. It is shown in this section that the mobility of hard spheres against concentration is the same independently of the simulation algorithm. Any of the three options, Glauber, Metropolis or interpolation, can be used for the description of the out-of-equilibrium behavior of hard spheres in a discrete lattice.

The hard-core interaction is given by
\begin{equation}\label{eq:hc}
\phi_{n_i} = \left\{ \begin{array}{cl}
0 & \text{if  } n_i=0, 1 \\ 
\infty & \text{if  } n_i \ge 2
\end{array}   \right. .
\end{equation}
Hard core imposes that, in any cell, $n_i$ takes values 0 or 1 only. 
Let us consider a jump from cell $i$ to cell $i+1$. The energy of the initial configuration $\{\cdots,n_i,n_{i+1},\cdots \}$ is $E_\text{ini}$, and the energy of the final configuration $\{\cdots,n_i-1,n_{i+1}+1,\cdots \}$ is $E_\text{fin}$. The external energy change is $\Delta U = -F a$, with $a$ the cell size. Then, the energy change is
\begin{align}
\Delta E &= E_\text{fin}-E_\text{ini} \nonumber\\
&= \Delta U + \phi_{n_i-1} + \phi_{n_{i+i}+1} - \phi_{n_i} - \phi_{n_{i+1}} \nonumber\\
&= \left\{ \begin{array}{cl}
-F a & \text{if  } n_{i+1}=0 \\ 
\infty & \text{if  } n_{i+1}=1
\end{array}   \right. ,
\label{eq:echange2}
\end{align}
where it was assumed that $n_i=1$ so that the jump process from $i$ to $i+1$ is possible (there must be a particle to make the jump).

Let us call $v$ the mean velocity of particles when a force $F$ is applied. Analytical expressions for the mobility, defined as $B=v/F$, can be derived for the three algorithms to compare with numerical results. Mobility is obtained from the mean current:
\begin{equation}\label{eq:current}
J = \langle n_i W_{i,i+1} - n_{i+1} W_{i+1,i}\rangle,
\end{equation}
since $J=v\, \bar{n}/a=B F \bar{n}/a$, so that $B = J\, a/(F\bar{n})$. We call the mobilities for Glauber and Metropolis algorithms $B^G$ and $B^M$ respectively. In the ideal system, where effects of interactions are neglected, the mobility is given by the Einstein relation: $B_0 = \beta D_0$, where $D_0=P a^2$ is the diffusivity.

First, for the Metropolis algorithm, the transition probability for a jump from $i$ to $i+1$, assuming that $F>0$, is
\begin{equation}\label{eq:methc}
W^M_{i,i+1} = P \min(1,e^{-\beta\, \Delta E}) = \left\{ \begin{array}{cl}
P & \text{if  } n_{i+1}=0 \\ 
0 & \text{if  } n_{i+1}=1
\end{array}   \right.,
\end{equation}
expression that can be written as
\begin{equation}\label{eq:methc2}
W^M_{i,i+1} = P (1-n_{i+1}).
\end{equation}
And, for the jump from $i+1$ to $i$,
\begin{equation}\label{eq:methc3}
W^M_{i+1,i} = P (1-\beta F a) (1-n_i),
\end{equation}
where it is assumed that the force and the cell size satisfy $Fa \ll k_B T$ so that $e^{-\beta F a} \simeq 1 - \beta F a$; this approximation is necessary to get a linear relation between mean velocity $v$ and force $F$, from which the mobility is obtained; the approximation corresponds to the linear regime of non equilibrium thermodynamics. Using \eqref{eq:methc2} and \eqref{eq:methc3} to evaluate the current \eqref{eq:current}, we get
\begin{equation}\label{eq:Jmethc}
J^M = P \beta F a\, \bar{n} (1 - \bar{n}),
\end{equation}
and the mobility is
\begin{equation}\label{eq:Bmethc}
\frac{B^M}{B_0} = 1 - \bar{n} \qquad \text{(hard core)}.
\end{equation}
It is convenient to present the mobility relative to $B_0$, so that the result is independent of the jump rate $P$ or the cell size $a$. In the last expressions, subindex $i$ is not used for the mean number of particles, $\bar{n}$, since a homogeneous system is considered.

For the Glauber algorithm, using the definition \eqref{eq:glauber}, it can be shown that
\begin{align}
W^G_{i,i+1} &= P (1+\beta F a/2) (1-n_{i+1}) \\
W^G_{i+1,i} &= P (1-\beta F a/2) (1-n_i).
\end{align}
Using \eqref{eq:current}, the result for the current $J^G$ turns out to be equal to $J^M$, and, of course, also the mobility:
\begin{equation}\label{eq:Bglahc}
\frac{B^G}{B_0} = 1 - \bar{n} \qquad \text{(hard core)}.
\end{equation}

For the interpolation algorithm, the first step is to calculate $V_i$. We have that
\begin{align}
e^{-\beta (\phi_{n_i+1}-\phi_{n_i})} &= \left\{ \begin{array}{cl}
1 & \text{if  } n_i=0 \\ 
0 & \text{if  } n_i=1
\end{array} \right. \nonumber \\
&= 1 - n_i,
\end{align}
then, from Eq.\ \eqref{eq:widom}, we have
\begin{equation}\label{eq:Vint}
V_i = - \beta^{-1} \ln(1-\bar{n}_i)
\end{equation}
and
\begin{equation}\label{eq:Vintp}
V_i' = \frac{\beta^{-1}}{1-\bar{n}_i}
\end{equation}
When replacing this expressions in Eq.\ \eqref{eq:transpr2} for the transition probability, as mentioned before, $V_i$ and $V_i'$ have to be evaluated at integer values of $n_i$. Then, using also that
\begin{equation}\label{eq:deltaV}
\Delta V = V_{i+1}-V_i = -\beta^{-1} \ln\left(\frac{1-n_{i+1}}{1-n_i}\right)
\end{equation}
it can be shown that the transition probabilities for the interpolation algorithm are equal to those of the Glauber algorithm:
\begin{align}
W^I_{i,i+1} &= P (1+\beta F a/2) (1-n_{i+1}) \\
W^I_{i+1,i} &= P (1-\beta F a/2) (1-n_i).
\end{align}
Therefore, the mobility is also the same,
\begin{equation}\label{eq:Binthc}
\frac{B^I}{B_0} = 1 - \bar{n} \qquad \text{(hard core)}.
\end{equation}

\begin{figure}
\includegraphics[width=\columnwidth]{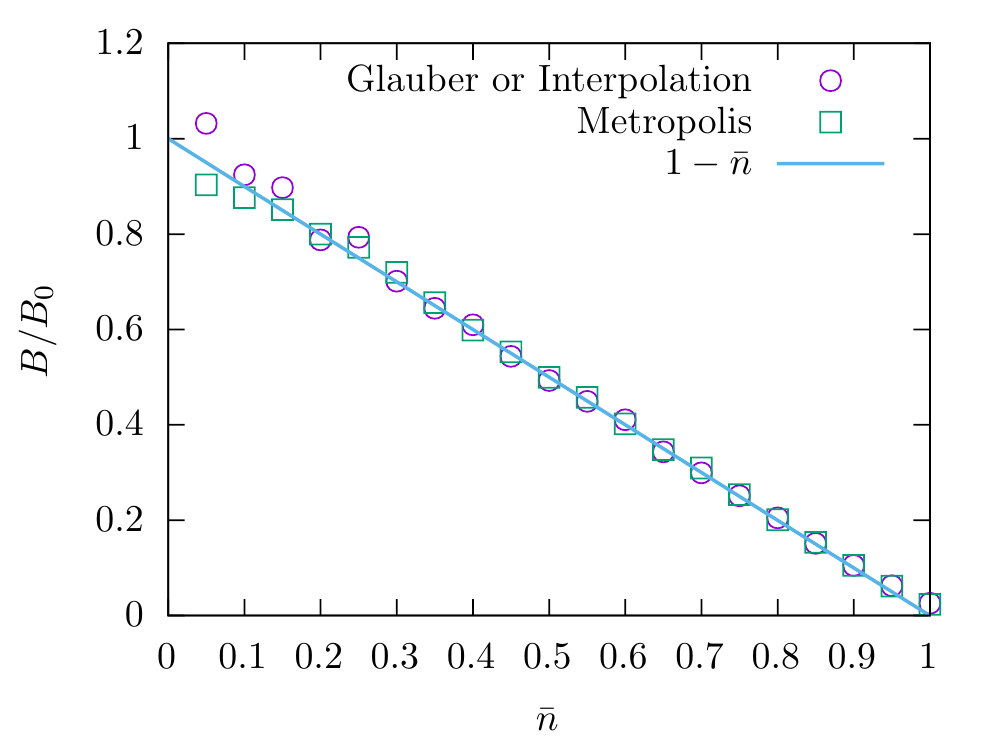}
\caption{Mobility $B/B_0$ against average number of particles per cell $\bar{n}$, using hard-core interaction in a lattice with $100$ cells, with periodic boundary conditions. Numerical results for Metropolis (squares) and Glauber or interpolation (circles) algorithms are in agreement with the prediction $B/B_0=1-\bar{n}$ (solid line). Number of Monte Carlo steps: $10^4$, applied force: $\beta F a = 0.05$.}
\label{fig:hc}
\end{figure}

Figure \ref{fig:hc} shows numerical results of the mobility using Metropolis and Glauber (or interpolation) algorithms for the hard-core interaction. The results match the expression obtained for $B/B_0$.

In summary, hard spheres are not useful to discriminate among algorithms. Any of them seems to work. This is not the case for the attractive interaction of a classical Bose-Einstein gas, and this is the main reason why we focus our attention on it.

\section{Basic formulae for the Bose-Einstein gas}
\label{sec:basic}

In the lattice gas description, the system is divided into cells of size $a$ much smaller than the characteristic length of particle density variations. Temperature $T$ and chemical potential $\mu$ are homogeneous. Interaction energy between particles in different cells is neglected. Since spatial variations are smooth, cells are point-like in the continuous limit, and local thermal equilibrium holds. This means that each cell can be considered as an equilibrium system, although the whole system is out of equilibrium. We can write the classical grand partition function of cell $i$ as
\begin{equation}\label{eq:partfunc}
\mathcal{Z}_i = \sum_{n_i=0}^{\infty} \frac{\exp[-\beta(\phi_{n_i}+n_i U_i -n_i \mu)]}{n_i!},
\end{equation}
where the energy in cell $i$ is $\phi_{n_i}+n_i U_i$.

We consider a configuration energy
\begin{equation}\label{eq:confen}
\phi_{n_i}= -\beta^{-1} \ln n_i!
\end{equation}
% in order
to reproduce Bose-Einstein statistics, since in this case we have
\begin{equation}
\mathcal{Z}_i = \sum_{n_i=0}^{\infty}\exp[-\beta n_i(U_i - \mu)]= \frac{1}{1-e^{-\beta (U_i-\mu)}},
\end{equation}
and the average number of particles is
\begin{equation}\label{eq:BEdist}
\bar{n}_i = \frac{1}{e^{\beta (U_i-\mu)}-1}.
\end{equation}
If we know that the system has a total number of particles $N$, the chemical potential is obtained from the relationship $N=\sum_i \bar{n}_i$, where the sum is performed in all cells. From Ec.~\eqref{eq:BEdist}, the chemical potential is
\begin{equation}\label{eq:chempot}
\mu = U_i+ \beta^{-1}\ln \bar{n}_i - \beta^{-1}\ln (1+\bar{n}_i).
\end{equation}
% where in the right hand side we have the combination of cell depending values of $U_i$ and $\bar{n}_i$ that gives the corresponding value of the chemical potential $\mu$.
Let us consider a cell in which the potential $U_i$ takes a value $\mu^\circ$, i.e. a reference chemical potential.
Then, in the previous expression, we can recognize the ideal and residual parts of the chemical potential:
\begin{equation}\label{eq:chempot2}
\mu = \underbrace{\mu^\circ + \beta^{-1}\ln \bar{n}_i}_\text{ideal} - \underbrace{\beta^{-1}\ln (1+\bar{n}_i)}_\text{residual}.
\end{equation}
%gives the chemical potential $\mu$ needed to have a given number of particles $\bar{n}_i$, and $\mu^\circ$ is the usual reference chemical potential.
%\begin{align}
%\mu_\text{id} &= \mu^\circ + \beta^{-1}\ln \bar{n}_i \\
%\tilde{\mu} &= - \beta^{-1}\ln (1+\bar{n}_i).
%\end{align}

This simplified description in terms of jumps between neighboring cells is intended to correctly reproduce the behavior of a system of particles which move with given velocities and interact with a space dependent potential. The effective potential, also known as statistical potential, between two bosons at distance $r$ is
\begin{equation}\label{eq:vstat}
v_s(r) = -\beta^{-1}\ln\left(1+e^{-2\pi r^2/\lambda^2}\right),
\end{equation}
where $\lambda$ determines the range of the interaction (it is equal to the de Broglie wavelength in the quantum case)~\cite[p.\ 138]{pathria}. Notice that here we use the term ``boson'' to informally refer to classical particles with Bose-Einstein statistics.
The statistical potential holds for small concentration, so we can expect descriptions to agree only in this limit. The concentration of a cell (in local equilibrium) is $\rho_i=\bar{n}_i/a^3$.
Also, in the limit of small concentration, we can use the cluster expansion to obtain the chemical potential:
\begin{equation}\label{eq:cluster}
\frac{e^{\beta \mu}}{\lambda^3} = \rho_i - b_2 \rho_i^2 + \mathcal{O}(\rho_i^3),
\end{equation}
% see for example Eq.\ (5.32) in Ref.~\cite{kardar};
where $b_2$ is the coefficient of two-particle clusters given by
\begin{equation}\label{eq:b2}
b_2 = \int d^3\mathbf{r}\, (e^{-\beta v_s(r)}-1) = \lambda^3/2^{3/2};
\end{equation}
see for example Eq.\ (5.32) in Ref.~\cite{kardar}.
The cluster expansion is based on the quantity $f(\mathbf{r}) = e^{-\beta v(r)}-1$ as a convenient expansion parameter, with $v(r)$ the interaction potential. For short-range hard-core interactions, it is equal to $-1$ for $r\rightarrow 0$ and decays to zero for increasing $r$. In our case, it takes the value $1$ for $r\rightarrow 0$ and vanishes exponentially as $r$ increases.

From Eqs.\ \eqref{eq:cluster} and \eqref{eq:b2} we have that the chemical potential is
\begin{equation}\label{eq:mu}
\mu \simeq \underbrace{\beta^{-1} \ln (\rho_i\lambda^3)}_\text{ideal} - \underbrace{\beta^{-1} \ln (1+\rho_i \lambda^3/2^{3/2})}_\text{residual},
\end{equation}
where, as in Eq.~\eqref{eq:chempot2}, we can identify the ideal and residual parts.

% It is possible to match both descriptions
By imposing the condition in which the chemical potential in the lattice gas description, Eq.\ \eqref{eq:chempot2}, is equal to the chemical potential in the kinetic description, Eq.\ \eqref{eq:mu},
% The relationship between cell size $a$ and interaction range $\lambda$ can now be set by imposing the condition in which the chemical potential in the lattice gas description, Eq.\ \eqref{eq:chempot2}, is equal to the chemical potential in the kinetic description, Eq.\ \eqref{eq:mu}.
more specifically,
% we obtain the sought information
by matching the residual parts of the chemical potential, we have that $\bar{n}_i = \rho_i\lambda^3/2^{3/2}$, and therefore,
\begin{equation}\label{eq:alambda}
a=\lambda/2^{1/2}.
\end{equation}
% Therefore, to compare both descriptions, the cell size $a$ can not be arbitrary; it is related to the interaction range $\lambda$ as in \eqref{eq:alambda}.
But setting this condition poses a problem in the calculation of the grand partition function, since in Eq.\ \eqref{eq:partfunc} interactions with particles in neighboring cells are neglected. This approximation is valid as long as $a$ is much larger than the interaction range, i.e. $a\gg\lambda$, a condition that is not fulfilled in \eqref{eq:alambda}. Still, the qualitative behavior of the chemical potential is the same in both descriptions. We expect equivalent qualitative behaviors also for the mobility although, due to the inconsistency of approximations, we can not expect quantitative agreement.

\section{Mobility from Glauber, Metropolis and interpolation algorithms}
\label{sec:gm}

The energy change for a jump from cell $i$ to $i+1$ with effective Bose-Einstein interaction is
\begin{align}
\Delta E &= \Delta U + \phi_{n_i-1} + \phi_{n_{i+i}+1} - \phi_{n_i} - \phi_{n_{i+1}} \nonumber\\
&= -Fa - \beta^{-1} \ln \frac{n_{i+1} + 1}{n_i},
\label{eq:echange}
\end{align}
where the condition $n_i \ge 1$ has to be fulfilled in order to have the possibility of a jump from cell $i$ (for cells with $n_i=0$ there is no jump process and no need to calculate $\Delta E$).

\begin{figure}
	\includegraphics[width=\columnwidth]{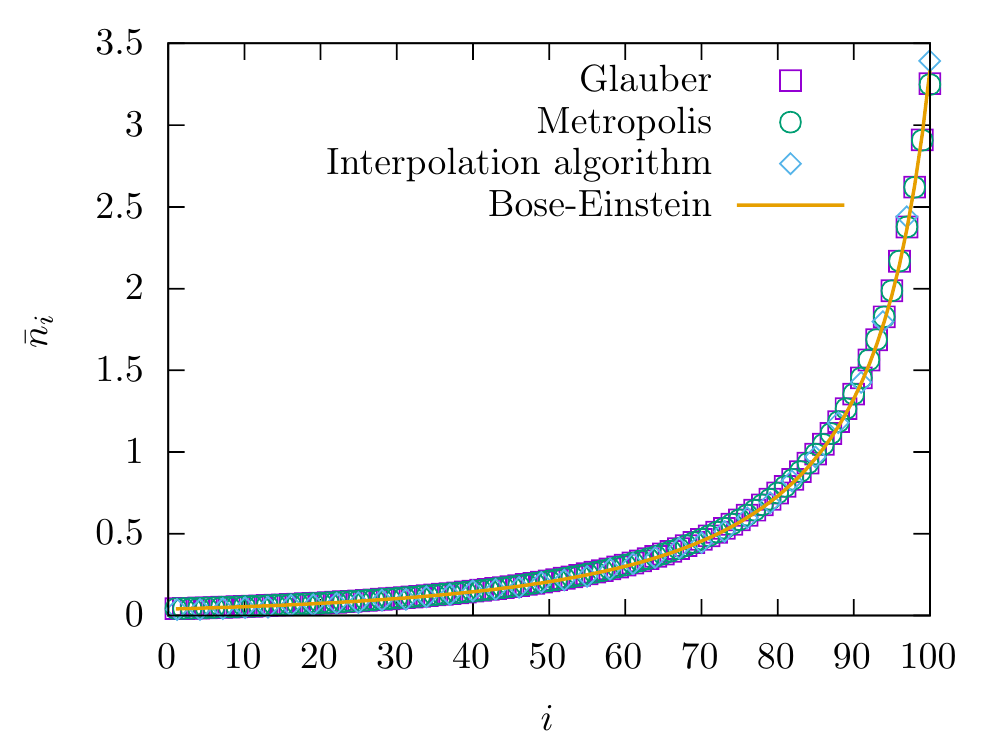}
	\caption{Equilibrium distribution of particles, $\bar{n}_i$, against position, $i$, for Glauber (squares), Metropolis (circles) and Interpolation (diamonds) algorithms. The interpolation algorithm data is from Ref.~\cite{suarez}.
	The curve corresponds to the Bose-Einstein distribution, Eq.\ \eqref{eq:BEdist} with $\beta U_i = -\beta F a i= - 0.03\, i$ and $\beta\mu$ calculated from the condition $N=\sum_i \bar{n}_i$ with $N$ the total number of particles. Number of Monte Carlo steps: $2\; 10^5$, lattice size: 100, number of particles $N=50$.}
	\label{fig:eq}
\end{figure}

With fixed boundary conditions and an external force, particles accumulate on one side of the system and, in equilibrium, they have the Bose-Einstein distribution given by Eq.~\eqref{eq:BEdist}. We have verified that Metropolis or Glauber algorithms converge to the Bose Einstein distribution in this case, as shown in Fig.\ \ref{fig:eq}. 

Now we turn to the non-equilibrium situation. Instead of fixed boundary conditions, let us consider periodic boundary conditions. After some time, the system is in a stationary non-equilibrium state. The mean velocity of a particle in terms of the transition probabilities between cells of size $a$ is
\begin{equation}
v=a W_{i,i+1} - a W_{i,i-1}.
\label{eq:vel}
\end{equation}

For small concentration (ideal gas), interactions can be neglected and $\Delta E \simeq -Fa$. It can be shown that for both algorithms, Glauber and Metropolis, the mean velocity for small concentration is $v=P a^2 \beta F$. For Metropolis at small concentration and a positive force $F$, transition probabilities are
\begin{equation*}
\begin{aligned}
W_{i,i+1}^M &= P \\
W_{i,i-1}^M &= P(1-\beta F a)
\end{aligned}  \qquad \text{(small concentration)}
\end{equation*}
Replacing in \eqref{eq:vel}, we obtain
\begin{equation*}
v^M = P a^2 \beta F \qquad\qquad \text{(small conc.)}.
\end{equation*}
For the Glauber algorithm we have
\begin{equation*}
\begin{aligned}
W_{i,i+1}^G &= \dfrac{2P}{1+e^{-\beta F a}}\simeq P(1+\beta F a/2) \\
W_{i,i-1}^G &= \dfrac{2P}{1+e^{\beta F a}}\simeq P(1-\beta F a/2).
\end{aligned}  \qquad \text{(small conc.)}
\end{equation*}
And, again, replacing in \eqref{eq:vel}, we obtain
\begin{equation*}
v^G = P a^2 \beta F \qquad\qquad \text{(small conc.)}.
\end{equation*}

The mobility, $B$, is defined as $v/F$. The previous results for the velocity, for both algorithms, are consistent with the Einstein relation; in both cases, $B_0 = \beta P a^2= \beta D_0$ is obtained. Subindex 0 identifies the small concentration regime.

As mentioned before, mobility is obtained from the mean current, Eq.\ \eqref{eq:current}. It can be shown that, for the Glauber algorithm,
\begin{align}\label{eq:BG}
\frac{B^G}{B_0} &= \frac{2}{\bar{n}} \sum_{n_i=0}^{\infty} \sum_{n_{i+1}=0}^{\infty} \frac{n_i^2(n_{i+1}+1) + n_{i+1}^2(n_i+1)}{(n_i+n_{i+1}+1)} \nonumber \\
& \qquad \times \frac{\bar{n}^{n_i + n_{i+1}}}{(1+\bar{n})^{n_i + n_{i+1}+2}},
\end{align}
and that the corresponding expression for the Metropolis algorithm can be simplified to
\begin{equation}\label{eq:BM}
\frac{B^M}{B_0} =\frac{1+\bar{n}}{1+2\bar{n}};
\end{equation}
see the Appendix for details.

In Fig.\ \ref{fig:gm}, we show the numerical results of the mobility obtained with both algorithms, compared with the analytical expressions \eqref{eq:BG} and \eqref{eq:BM}. The agreement between numerical and analytical results supports their validity. Both algorithms coincide in the limit of small concentration, where $B=B_0$ and interactions can be neglected. But as the mean number of particles is increased, discrepancies grow. There is a quantitative difference of about 30\% between both predictions of $B$ for larger $\bar{n}$. It is well known that we can not expect a correct description of a non-equilibrium state with these algorithms. Nevertheless, it is interesting to determine how far from the correct result they are. This is the purpose of the next paragraphs.

%In next sections we show that there is a qualitative difference: instead of decreasing with $\bar{n}$, the relative mobility $B/B_0$ should increase for the boson effective interaction.

\begin{figure}
\includegraphics[width=\columnwidth]{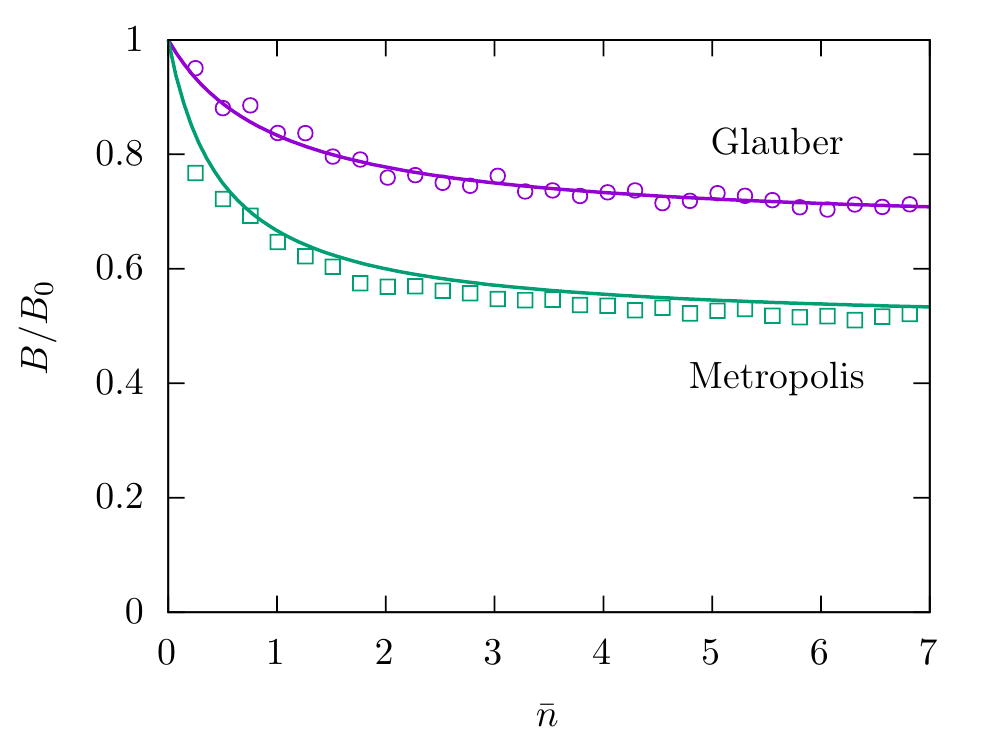}
\caption{Mobility $B/B_0$ for a system of particles with bosonic interaction obtained with Glauber and Metropolis algorithms, as a function of the average number of particles per cell $\bar{n}$.
Points represent numerical results while curves correspond to the analytical expressions \eqref{eq:BG} and \eqref{eq:BM} for the Glauber and Metropolis algorithms respectively. Each sum in Eq.\ \eqref{eq:BG} was approximated using a maximum number of terms equal to $10$. Numerical results were obtained using a lattice of 100 cells, during $10^7$ Monte Carlo steps, with an applied force $\beta F a =0.03$.}
\label{fig:gm}
\end{figure}

%\section{Mobility from interpolation algorithm}
%\label{sec:inter}

\begin{figure}
	\includegraphics[width=\columnwidth]{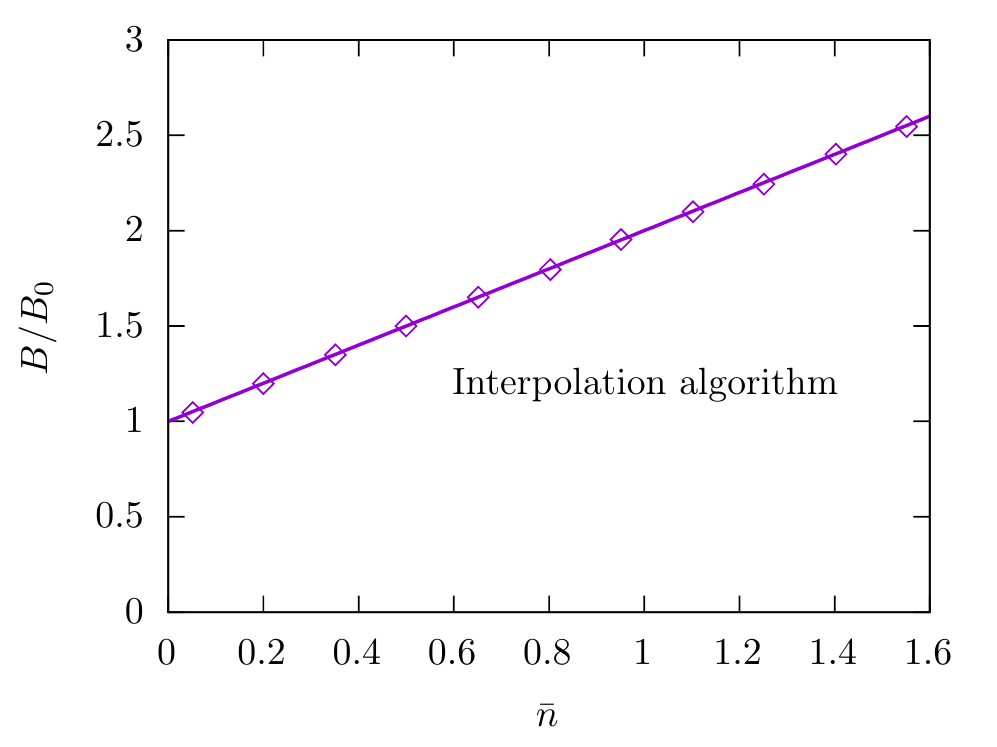}
	\caption{Mobility $B/B_0$ for a system of particles with bosonic interaction obtained with the interpolation algorithm, as a function of the average number of particles per cell $\bar{n}$.
	Symbols $\diamond$ represent numerical results taken from Ref.\ \cite{suarez} (applied force $\beta F a =0.05$), while the line corresponds to Eq.~\eqref{eq:mobinter}.} %Numerical results were obtained using a lattice of 100 cells, during $10^6$ Monte Carlo steps, 5000 samples.
	\label{Binterp}
\end{figure}

For the interpolation algorithm, using Eq.\ \eqref{eq:confen} for $\phi_{n_i}$ in \eqref{eq:widom}, we have
\begin{equation}\label{eq:V}
V_i=-\beta^{-1} \ln (\bar{n}_i+1).
\end{equation}
Notice that $V_i$ is equal to the residual part of the chemical potential, see Eq.~\eqref{eq:chempot2}, and Eq.~\eqref{eq:widom} corresponds to the Widom insertion method, where $\phi_{n_i+1}-\phi_{n_i}$ is the insertion energy, i.e. the interaction energy needed to insert one particle. Using this result for $V_i$, and its derivative, in \eqref{eq:transpr2}, the transition probability is
\begin{equation}\label{eq:transinter}
W_{i,i+1} = P \, e^{-\beta\,\Delta U/2} (1+n_{i+1}).
\end{equation}
With this information, we can calculate the mobility $B^I$ for the interpolation algorithm (see the Appendix). The result is
\begin{equation}\label{eq:mobinter}
\frac{B^I}{B_0} = 1 + \bar{n}.
\end{equation}
As in the previous results, a homogeneous system, in which $\bar{n}_i = \bar{n}_{i+1}=\bar{n}$, is considered for the calculation of mobility.
Figure \ref{Binterp} shows a good agreement between this theoretical result and kinetic Monte Carlo simulations.

The result for the mobility obtained with the interpolation algorithm is qualitatively different from the results of Glauber and Metropolis algorithms. While for the interpolation algorithm we have a mobility which increases with $\bar{n}$, for the other algorithms it decreases (see Fig.\ \ref{fig:gm}). As mentioned before, we know that Glauber and Metropolis algorithms are designed to give the correct equilibrium state; they should not be applied to non-equilibrium situations, but it is interesting to evaluate the error. According to the interpolation algorithm, which \textit{is} designed for non-equilibrium states (with limitations that are summarized in the conclusions), the error increases with concentration. Glauber and Metropolis algorithms give the correct result for the mobility only in the limit of small $\bar{n}$, where interactions can be neglected.

%Now, we can verify the validity of the interpolation algorithm using molecular dynamics.

\section{Molecular dynamics}
\label{sec:md}

The objective of this section is to obtain the mobility of a boson gas in the context of a classical kinetic description which includes velocity of particles, and compare with the results of the previous sections.

The method is to numerically obtain the self-diffusivity $D$ and use the Einstein relation to calculate the mobility $B=\beta D$. The Green-Kubo formula (see for example \cite[Sec.\ 4.6.2]{kubo} or \cite[Sec.\ S10.G]{reichl}) is used to obtain the self-diffusivity from the velocity autocorrelation function:
\begin{equation}\label{eq:selfdif}
D = \frac{1}{3}\int_{0}^{\infty} d t'\, \langle\mathbf{v}(0)\cdot\mathbf{v}(t')\rangle.
\end{equation}
We wish to emphasize that with this method the mobility is obtained from simulations of the system in equilibrium, without a force applied. The main purpose of this paper is to check algorithms for Monte Carlo simulations out of equilibrium. The motivation of this section is to obtain mobility using molecular dynamics, and in this context we can choose the method based on simplicity.

To compare with results of the previous sections, we have to calculate the mobility $B$ relative to its value for the ideal gas, $B_0$. From the kinetic theory of transport in dilute gases (see for example \cite[Sec.\ 16-1]{mcquarrie}), we know that the self-diffusivity in the ideal case, and therefore $B_0$, behaves as $1/\rho$, where $\rho$ is the particle density. The proportionality constant, between $B_0$ and $1/\rho$, is numerically set in our results so as to have $B/B_0 \rightarrow 1$ for $\rho\rightarrow 0$.

Then, we perform molecular dynamic simulations of a system of particles with a given density $\rho$, in equilibrium, which interact among them with the statistical potential of Eq.\ \eqref{eq:vstat}.
We do this for different densities and obtain the mobility through the velocity autocorrelation function. We focus our attention on the slope of the mobility for small concentrations, since this is the limit for which the statistical potential \eqref{eq:vstat} holds.
% \textbf{Also, $\lambda/a<3\%$, which meets the lattice gas description approximation, i.e. $a$ much larger than the interaction range. (no entiendo de donde sale esa relación entre $a$ y $\lambda$)}
% To compare with the results of the previous sections, we use an adimensional density $\rho^* = \rho a^3=\rho \lambda^3/2^{3/2}$ (see Eq.~\eqref{eq:alambda}).

\begin{figure}
	\includegraphics[width=\columnwidth]{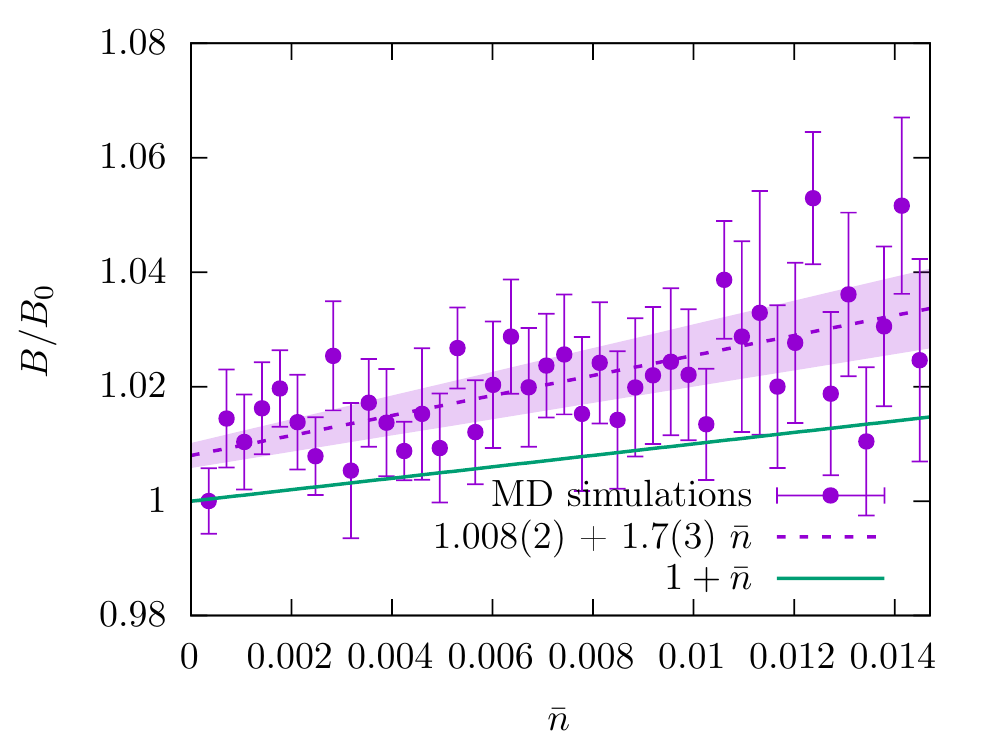}
	\caption{Mobility $B/B_0$ against $\bar{n}=\rho \lambda^3/2^{3/2}$ obtained from molecular dynamics simulations.
	We use LAMMPS software~\cite{lammps,plimpton1993fast} with parameters:
	temperature $T=1$, Boltzmann constant $k_B=1$, number of particles $2000$, cut-off distance $r_c=2.5$, $\lambda=1$,
	nve integration method and $10^7$ run steps.
	The interactions among particles are given by the statistical potential for bosons, Eq.~\eqref{eq:vstat} (this defines a new pair style in LAMMPS software). For each simulation, we obtain a value of $B$ for a given $\rho$. We perform around $90$ realizations for the same $\rho$ (with different initial conditions) to average.
	The dashed line corresponds to a linear fit ($1.008 + 1.7\bar{n}$), while the solid area represents its error, i.e. the area between $1.006 + 1.4\,\bar{n}$ and $1.010 + 2.0\,\bar{n}$. 
	We added the linear curve corresponding to the interpolation algorithm for comparison (solid line); see  Eq.~\eqref{eq:mobinter}.
% 	\textbf{eliminar el asterisco en $\bar{n}^*$, y agregar la recta del algoritmo de interpolación}
% 	whose behavior agrees with Eq.~\eqref{eq:mobinter}.
	}
	\label{fig:md}
\end{figure}

In Fig. \ref{fig:md}, we show the data obtained for $B$  as a function of $\bar{n}=\rho \lambda^3/2^{3/2}$.
% (\textbf{hay que modificar este valor, otra alternativa es mostrar rho*lambda$^3$ en lugar de rho*a$^3$, para mostrar comportamiento cualitativo sería lo mismo, pero creo que es mejor con a$^3$ y mostrar una pendiente distinta de 1})
This result supports the validity of the interpolation algorithm since the same qualitative behavior is obtained: an approximate linear increase of mobility with concentration with slopes of the same order, see Fig.\ \ref{fig:md}.

\section{Conclusions}
\label{sec:conclusions}

We study mobility in a system of classical particles with interactions. Two different approaches are possible: a lattice gas with transition probabilities among cells, or a kinetic description in a continuous space. A fundamental problem of the lattice gas description is to determine transition probabilities when only state energies are known. Monte Carlo simulations with Glauber or Metropolis algorithms correctly describe the equilibrium state but they are not supposed to hold out of equilibrium. Instead, the interpolation algorithm was recently introduced \cite{suarez,mdp,dipietro2,dipietro3} in order to obtain transition probabilities that hold out of equilibrium; it is summarized in Eqs. \eqref{eq:widom} and \eqref{eq:transpr2}. Limitations of the method are: the system should be in local thermal equilibrium (i.e. deviations from equilibrium have a characteristic length much larger than the cell size), no phase transition occur and the interaction is local [more precisely, the interaction energy can be written as in \eqref{eq:energy}]. Also, the information provided by the algorithm is incomplete if the jump rate $P$ has a non-trivial dependence on concentration, as it happens, for example, for diffusion in a solid \cite{dipietro3}.

We calculate the mobility in a non-equilibrium stationary state: a force is applied to a one-dimensional array of cells with periodic boundary conditions and homogeneous density. For hard core interaction, the same results are obtained for Glauber, Metropolis or interpolation algorithms. Instead, for an attractive potential such that the Bose-Einstein statistics is reproduced, important differences are obtained among algorithms. There is a difference of about 30\% between the Glauber and Metropolis predictions for large concentration; in both cases, mobility decreases with concentration towards an asymptotic value.
The mobility obtained from the interpolation algorithm qualitatively differs from the ones of the other methods. Instead of decreasing with concentration, it increases. This means an unbounded increasing error for the Glauber and Metropolis algorithms. They can be used to calculate mobility only in the limit of small concentration, where $B\rightarrow B_0$.

The lattice gas description should be consistent with the kinetic description. So, we also consider particles moving in a continuous space and interacting with a statistical potential \eqref{eq:vstat} which corresponds to the Bose-Einstein distribution. The mobility obtained from molecular dynamics simulations, Fig.\ \ref{fig:md}, is in qualitative agreement with the prediction of the interpolation algorithm.

\begin{acknowledgments}
The authors acknowledge discussions with H. M\'artin which were useful for the development of these ideas. This work was partially supported by Consejo Nacional de Investigaciones Cient\'ificas y T\'ecnicas (CONICET, Argentina, PIP 112 201501 00021 CO).
This work used computational resources from CCAD -- Universidad Nacional de C\'ordoba~\cite{ccad}, in particular the Mendieta Cluster, which is part of SNCAD -- MinCyT, Rep\'ublica Argentina.
\end{acknowledgments}

\section*{Appendix}

In this appendix, the expressions \eqref{eq:BG} and \eqref{eq:BM} for the mobility for Glauber and Metropolis algorithms are derived.

Using \eqref{eq:echange} for the energy change, the transition probabilities for the Glauber algorithm are
\begin{align}
W_{i,i+1} &= 2P\frac{n_{i+1}+1}{n_i e^{-\beta F a} + n_{i+1}+1}, \\
W_{i+1,i} &= 2P\frac{n_i+1}{n_{i+1} e^{\beta F a} + n_i+1}.
\end{align}
Replacing these expressions in the current $J=\langle \mathcal{J}\rangle$, with
\begin{equation}\label{eq:current2}
\mathcal{J} = n_i W_{i,i+1} - n_{i+1} W_{i+1,i},
\end{equation}
and making approximations for $\beta F a \ll 1$, we have
\begin{align}
\frac{J}{2P}&=\left\langle \frac{n_i-n_{i+1}}{n_i+n_{i+1}+1}\right\rangle \nonumber\\
& \quad + \left\langle\frac{\beta F a( n_i^2(n_{i+1}+1) + n_{i+1}^2(n_i+1))}{(n_i+n_{i+1}+1)^2} \right\rangle.
\end{align}
The first term of the right side cancels due to the symmetry of the homogeneous stationary state. Knowing that $B=J\, a/(F\bar{n})$ and that $B_0=\beta P a^2$, we have
\begin{equation}
\frac{B^G}{B_0} = \frac{J}{\bar{n}P \beta F a}=\frac{2}{\bar{n}}\left\langle\frac{ n_i^2(n_{i+1}+1) + n_{i+1}^2(n_i+1)}{(n_i+n_{i+1}+1)^2} \right\rangle.
\end{equation}
The probability of having $n_i$ particles knowing that the average value is $\bar{n}$, for the Bose-Einstein distribution, is given by
\begin{equation}\label{eq:pBE}
p_\text{BE}(n_i) = \frac{\bar{n}^{n_i}}{(\bar{n}+1)^{n_i + 1}};
\end{equation}
see for example \cite[p.\ 152]{pathria}.
Using this probability, we obtain Eq.\ \eqref{eq:BG} for the mobility in the Glauber algorithm. Actually, a correction of order $\beta F a$ should be added to $p_\text{BE}(n_i)$ to use the probability which corresponds to the stationary non-equilibrium state, since Eq.\ \eqref{eq:pBE} holds in equilibrium. But this correction cancels when only terms up to order $\beta F a$ are kept in the equation for the current.

A similar process is applied to obtain the mobility for the Metropolis algorithm. Replacing the expression \eqref{eq:echange} for $\Delta E$ in the Metropolis transition probabilities, we have
\begin{align}
\frac{\mathcal{J}}{P}&=n_i \min\left( 1, \frac{n_{i+1}+1}{n_i}e^{\beta F a} \right)\nonumber \\
&\quad - n_{i+1} \min\left( 1, \frac{n_i + 1}{n_{i+1}}e^{-\beta F a} \right).
\end{align}
Assuming that $\beta F a \ll 1$ and considering all possible combinations of $n_i$ and $n_{i+1}$, we get
\begin{equation}
\frac{\mathcal{J}}{P}=\left\{ \begin{array}{cl}
 (n_i + 1)\beta F a-1 & \text{ if } n_{i+1}\ge n_i + 1 \\
0 & \text{ if } n_{i+1} = n_i  \\
1 & \text{ if } n_{i+1} = n_i-1 \\
1 + (n_{i+1} + 1)\beta F a & \text{ if } n_{i+1} \le n_i-2
\end{array}  \right. .
\end{equation}
The average of this expression can be written as
\begin{align}
\frac{J}{P} &= \sum_{n_i=2}^{\infty} \sum_{n_{i+1}=0}^{n_i-2} [1+(n_{i+1}+1)\beta F a] p_\text{BE}(n_{i+1})p_\text{BE}(n_i) \nonumber\\
&\ \ + \sum_{n_i=1}^{\infty} p_\text{BE}(n_i-1) p_\text{BE}(n_i) \nonumber \\
&\ \ + \sum_{n_i=0}^{\infty} \sum_{n_{i+1}=n_i+1}^{\infty} [(n_i+1)\beta F a-1]p_\text{BE}(n_{i+1})p_\text{BE}(n_i).
\end{align}
These sums can be simplified. Replacing $p_\text{BE}(n_i)=q^{n_i}/(\bar{n}+1)$ with $q=\bar{n}/(\bar{n}+1)$, and using a symbolic manipulator such as Maxima~\cite{maxima}, we obtain
\begin{align}
\frac{J}{P} &= \frac{q\beta F a}{(1-q)^2 (1-q) (1+q) (1+\bar{n})^2} \nonumber \\
&= \frac{\bar{n}(1+\bar{n})}{1+2\bar{n}}\beta F a.
\end{align}
Using the relationship between current and mobility, $B=J\, a/(F\bar{n})$, Eq.\ \eqref{eq:BM} is immediately obtained.

Finally, the calculation of the mobility for the interpolation algorithm is simpler. We have the transition probabilities in \eqref{eq:transinter}, and the mean current is
\begin{align}
J &= P \langle n_i (1 + n_{i+1}) e^{\beta F a/2} - n_{i+1}(1+n_i) e^{-\beta F a/2} \nonumber \\
&\simeq P \langle n_i n_{i+1} \beta F a + n_i(1 + \beta F a/2) - n_{i+1} (1 - \beta F a/2) \rangle \nonumber\\
&= P \beta F a \bar{n}(1+\bar{n}),
\end{align}
where it was considered that $\langle n_i\rangle=\langle n_{i+1}\rangle =\bar{n}$ and that fluctuations in different cells are independent, so $\langle n_i n_{i+1}\rangle = \bar{n}^2$.
From this equation for $J$, the result for the mobility \eqref{eq:mobinter} is obtained.

\bibliography{bosons.bib}

\begin{thebibliography}{23}%
\makeatletter
\providecommand \@ifxundefined [1]{%
 \@ifx{#1\undefined}
}%
\providecommand \@ifnum [1]{%
 \ifnum #1\expandafter \@firstoftwo
 \else \expandafter \@secondoftwo
 \fi
}%
\providecommand \@ifx [1]{%
 \ifx #1\expandafter \@firstoftwo
 \else \expandafter \@secondoftwo
 \fi
}%
\providecommand \natexlab [1]{#1}%
\providecommand \enquote  [1]{``#1''}%
\providecommand \bibnamefont  [1]{#1}%
\providecommand \bibfnamefont [1]{#1}%
\providecommand \citenamefont [1]{#1}%
\providecommand \href@noop [0]{\@secondoftwo}%
\providecommand \href [0]{\begingroup \@sanitize@url \@href}%
\providecommand \@href[1]{\@@startlink{#1}\@@href}%
\providecommand \@@href[1]{\endgroup#1\@@endlink}%
\providecommand \@sanitize@url [0]{\catcode `\\12\catcode `\$12\catcode
  `\&12\catcode `\#12\catcode `\^12\catcode `\_12\catcode `\%12\relax}%
\providecommand \@@startlink[1]{}%
\providecommand \@@endlink[0]{}%
\providecommand \url  [0]{\begingroup\@sanitize@url \@url }%
\providecommand \@url [1]{\endgroup\@href {#1}{\urlprefix }}%
\providecommand \urlprefix  [0]{URL }%
\providecommand \Eprint [0]{\href }%
\providecommand \doibase [0]{http://dx.doi.org/}%
\providecommand \selectlanguage [0]{\@gobble}%
\providecommand \bibinfo  [0]{\@secondoftwo}%
\providecommand \bibfield  [0]{\@secondoftwo}%
\providecommand \translation [1]{[#1]}%
\providecommand \BibitemOpen [0]{}%
\providecommand \bibitemStop [0]{}%
\providecommand \bibitemNoStop [0]{.\EOS\space}%
\providecommand \EOS [0]{\spacefactor3000\relax}%
\providecommand \BibitemShut  [1]{\csname bibitem#1\endcsname}%
\let\auto@bib@innerbib\@empty
%</preamble>
\bibitem [{\citenamefont {Bortz}\ \emph {et~al.}(1975)\citenamefont {Bortz},
  \citenamefont {Kalos},\ and\ \citenamefont {Lebowitz}}]{bortz}%
  \BibitemOpen
  \bibfield  {author} {\bibinfo {author} {\bibfnamefont {A.~B.}\ \bibnamefont
  {Bortz}}, \bibinfo {author} {\bibfnamefont {M.~H.}\ \bibnamefont {Kalos}}, \
  and\ \bibinfo {author} {\bibfnamefont {J.~L.}\ \bibnamefont {Lebowitz}},\
  }\bibfield  {title} {\enquote {\bibinfo {title} {A new algorithm for {M}onte
  {C}arlo simulation of {I}sing spin systems},}\ }\href@noop {} {\bibfield
  {journal} {\bibinfo  {journal} {J. Comp. Phys.}\ }\textbf {\bibinfo {volume}
  {17}},\ \bibinfo {pages} {10} (\bibinfo {year} {1975})}\BibitemShut {NoStop}%
\bibitem [{\citenamefont {Fichthorn}\ and\ \citenamefont
  {Weinberg}(1991)}]{fichthorn}%
  \BibitemOpen
  \bibfield  {author} {\bibinfo {author} {\bibfnamefont {K.~A.}\ \bibnamefont
  {Fichthorn}}\ and\ \bibinfo {author} {\bibfnamefont {W.~H.}\ \bibnamefont
  {Weinberg}},\ }\bibfield  {title} {\enquote {\bibinfo {title} {Theoretical
  foundations of dynamical {M}onte {C}arlo simulations},}\ }\href@noop {}
  {\bibfield  {journal} {\bibinfo  {journal} {J. Chem. Phys.}\ }\textbf
  {\bibinfo {volume} {95}},\ \bibinfo {pages} {1090} (\bibinfo {year}
  {1991})}\BibitemShut {NoStop}%
\bibitem [{\citenamefont {Voter}(2005)}]{voter}%
  \BibitemOpen
  \bibfield  {author} {\bibinfo {author} {\bibfnamefont {A.~F.}\ \bibnamefont
  {Voter}},\ }\bibfield  {title} {\enquote {\bibinfo {title} {Introduction to
  the kinetic monte carlo method},}\ }in\ \href@noop {} {\emph {\bibinfo
  {booktitle} {Radiation Effects in Solids}}},\ \bibinfo {editor} {edited by\
  \bibinfo {editor} {\bibfnamefont {K.~E.}\ \bibnamefont {Sickafus}}\ and\
  \bibinfo {editor} {\bibfnamefont {E.~A.}\ \bibnamefont {Kotomin}}}\ (\bibinfo
   {publisher} {Springer},\ \bibinfo {year} {2005})\BibitemShut {NoStop}%
\bibitem [{\citenamefont {Jansen}(2012)}]{jansen}%
  \BibitemOpen
  \bibfield  {author} {\bibinfo {author} {\bibfnamefont {A.~P.~J.}\
  \bibnamefont {Jansen}},\ }\href@noop {} {\emph {\bibinfo {title} {An
  Introduction to Kinetic {M}onte {C}arlo Simulations of Surface Reactions}}}\
  (\bibinfo  {publisher} {Springer},\ \bibinfo {address} {Berlin, Heidelberg},\
  \bibinfo {year} {2012})\BibitemShut {NoStop}%
\bibitem [{\citenamefont {Binder}(1997)}]{binder}%
  \BibitemOpen
  \bibfield  {author} {\bibinfo {author} {\bibfnamefont {K.}~\bibnamefont
  {Binder}},\ }\bibfield  {title} {\enquote {\bibinfo {title} {Applications of
  {M}onte {C}arlo methods to statistical physics},}\ }\href@noop {} {\bibfield
  {journal} {\bibinfo  {journal} {Rep. Prog. Phys.}\ }\textbf {\bibinfo
  {volume} {60}},\ \bibinfo {pages} {487} (\bibinfo {year} {1997})}\BibitemShut
  {NoStop}%
\bibitem [{\citenamefont {Binder}\ and\ \citenamefont
  {Heermann}(2010)}]{binder2}%
  \BibitemOpen
  \bibfield  {author} {\bibinfo {author} {\bibfnamefont {K.}~\bibnamefont
  {Binder}}\ and\ \bibinfo {author} {\bibfnamefont {D.~W.}\ \bibnamefont
  {Heermann}},\ }\href@noop {} {\emph {\bibinfo {title} {{M}onte {C}arlo
  Simulation in Statistical Physics}}}\ (\bibinfo  {publisher} {Springer},\
  \bibinfo {address} {Berlin, Heidelberg},\ \bibinfo {year} {2010})\BibitemShut
  {NoStop}%
\bibitem [{\citenamefont {Su\'arez}\ \emph {et~al.}(2015)\citenamefont
  {Su\'arez}, \citenamefont {Hoyuelos},\ and\ \citenamefont
  {M\'artin}}]{suarez}%
  \BibitemOpen
  \bibfield  {author} {\bibinfo {author} {\bibfnamefont {G.}~\bibnamefont
  {Su\'arez}}, \bibinfo {author} {\bibfnamefont {M.}~\bibnamefont {Hoyuelos}},
  \ and\ \bibinfo {author} {\bibfnamefont {H.}~\bibnamefont {M\'artin}},\
  }\bibfield  {title} {\enquote {\bibinfo {title} {Mean-field approach for
  diffusion of interacting particles},}\ }\href {\doibase
  10.1103/PhysRevE.92.062118} {\bibfield  {journal} {\bibinfo  {journal} {Phys.
  Rev. E}\ }\textbf {\bibinfo {volume} {92}},\ \bibinfo {pages} {062118}
  (\bibinfo {year} {2015})}\BibitemShut {NoStop}%
\bibitem [{\citenamefont {Mart\'{\i}nez}\ and\ \citenamefont
  {Hoyuelos}(2018)}]{mdp}%
  \BibitemOpen
  \bibfield  {author} {\bibinfo {author} {\bibfnamefont {M.~Di~Pietro}\
  \bibnamefont {Mart\'{\i}nez}}\ and\ \bibinfo {author} {\bibfnamefont
  {M.}~\bibnamefont {Hoyuelos}},\ }\bibfield  {title} {\enquote {\bibinfo
  {title} {Mean-field approach to diffusion with interaction: Darken equation
  and numerical validation},}\ }\href {\doibase 10.1103/PhysRevE.98.022121}
  {\bibfield  {journal} {\bibinfo  {journal} {Phys. Rev. E}\ }\textbf {\bibinfo
  {volume} {98}},\ \bibinfo {pages} {022121} (\bibinfo {year}
  {2018})}\BibitemShut {NoStop}%
\bibitem [{\citenamefont {Mart\'{\i}nez}\ and\ \citenamefont
  {Hoyuelos}(2019{\natexlab{a}})}]{dipietro2}%
  \BibitemOpen
  \bibfield  {author} {\bibinfo {author} {\bibfnamefont {M.~Di~Pietro}\
  \bibnamefont {Mart\'{\i}nez}}\ and\ \bibinfo {author} {\bibfnamefont
  {M.}~\bibnamefont {Hoyuelos}},\ }\bibfield  {title} {\enquote {\bibinfo
  {title} {From diffusion experiments to mean-field theory simulations and
  back},}\ }\href@noop {} {\bibfield  {journal} {\bibinfo  {journal} {J. Stat.
  Mech.: Theory Exp.}\ }\textbf {\bibinfo {volume} {2019}},\ \bibinfo {pages}
  {113201} (\bibinfo {year} {2019}{\natexlab{a}})}\BibitemShut {NoStop}%
\bibitem [{sup()}]{supmat}%
  \BibitemOpen
  \href@noop {} {}\bibinfo {note} {See Supplemental Material at \url{site},
  which includes
  Refs.~\cite{difcode,lammps,plimpton1993fast,lammpsdoc,pathria,mdcode}, for
  the details about the numerical implementation and the codes used to obtain
  the mobility results shown in this article.}\BibitemShut {Stop}%
\bibitem [{\citenamefont {Glauber}(1963)}]{glauber}%
  \BibitemOpen
  \bibfield  {author} {\bibinfo {author} {\bibfnamefont {R.~J.}\ \bibnamefont
  {Glauber}},\ }\bibfield  {title} {\enquote {\bibinfo {title}
  {Time‐dependent statistics of the {I}sing model},}\ }\href@noop {}
  {\bibfield  {journal} {\bibinfo  {journal} {J. Math. Phys.}\ }\textbf
  {\bibinfo {volume} {4}},\ \bibinfo {pages} {294} (\bibinfo {year}
  {1963})}\BibitemShut {NoStop}%
\bibitem [{\citenamefont {Hastings}(1970)}]{hastings}%
  \BibitemOpen
  \bibfield  {author} {\bibinfo {author} {\bibfnamefont {W.~K.}\ \bibnamefont
  {Hastings}},\ }\bibfield  {title} {\enquote {\bibinfo {title} {Monte {C}arlo
  sampling methods using {M}arkov chains and their applications},}\ }\href@noop
  {} {\bibfield  {journal} {\bibinfo  {journal} {Biometrika}\ }\textbf
  {\bibinfo {volume} {57}},\ \bibinfo {pages} {97} (\bibinfo {year}
  {1970})}\BibitemShut {NoStop}%
\bibitem [{\citenamefont {Vattulainen}\ \emph {et~al.}(1998)\citenamefont
  {Vattulainen}, \citenamefont {Merikoski}, \citenamefont {Ala-Nissila},\ and\
  \citenamefont {Ying}}]{vattulainen}%
  \BibitemOpen
  \bibfield  {author} {\bibinfo {author} {\bibfnamefont {I.}~\bibnamefont
  {Vattulainen}}, \bibinfo {author} {\bibfnamefont {J.}~\bibnamefont
  {Merikoski}}, \bibinfo {author} {\bibfnamefont {T.}~\bibnamefont
  {Ala-Nissila}}, \ and\ \bibinfo {author} {\bibfnamefont {S.~C.}\ \bibnamefont
  {Ying}},\ }\bibfield  {title} {\enquote {\bibinfo {title} {Adatom dynamics
  and diffusion in a model of {O/W}(110)},}\ }\href@noop {} {\bibfield
  {journal} {\bibinfo  {journal} {Phys. Rev. B}\ }\textbf {\bibinfo {volume}
  {57}},\ \bibinfo {pages} {1896} (\bibinfo {year} {1998})}\BibitemShut
  {NoStop}%
\bibitem [{\citenamefont {Mart\'{\i}nez}\ and\ \citenamefont
  {Hoyuelos}(2019{\natexlab{b}})}]{dipietro3}%
  \BibitemOpen
  \bibfield  {author} {\bibinfo {author} {\bibfnamefont {M.~Di~Pietro}\
  \bibnamefont {Mart\'{\i}nez}}\ and\ \bibinfo {author} {\bibfnamefont
  {M.}~\bibnamefont {Hoyuelos}},\ }\bibfield  {title} {\enquote {\bibinfo
  {title} {Diffusion in binary mixtures: an analysis of the dependence on the
  thermodynamic factor},}\ }\href@noop {} {\bibfield  {journal} {\bibinfo
  {journal} {Phys. Rev. E}\ }\textbf {\bibinfo {volume} {100}},\ \bibinfo
  {pages} {022112} (\bibinfo {year} {2019}{\natexlab{b}})}\BibitemShut
  {NoStop}%
\bibitem [{\citenamefont {Pathria}\ and\ \citenamefont
  {Beale}(2011)}]{pathria}%
  \BibitemOpen
  \bibfield  {author} {\bibinfo {author} {\bibfnamefont {R.~K.}\ \bibnamefont
  {Pathria}}\ and\ \bibinfo {author} {\bibfnamefont {P.~D.}\ \bibnamefont
  {Beale}},\ }\href@noop {} {\emph {\bibinfo {title} {Statistical
  Mechanics}}},\ \bibinfo {edition} {3rd}\ ed.\ (\bibinfo  {publisher}
  {Elsevier},\ \bibinfo {year} {2011})\BibitemShut {NoStop}%
\bibitem [{\citenamefont {Kardar}(2007)}]{kardar}%
  \BibitemOpen
  \bibfield  {author} {\bibinfo {author} {\bibfnamefont {M.}~\bibnamefont
  {Kardar}},\ }\href@noop {} {\emph {\bibinfo {title} {Statistical Physics of
  Particles}}}\ (\bibinfo  {publisher} {Cambridge University Press},\ \bibinfo
  {address} {Cambridge},\ \bibinfo {year} {2007})\BibitemShut {NoStop}%
\bibitem [{\citenamefont {Kubo}\ \emph {et~al.}(1998)\citenamefont {Kubo},
  \citenamefont {Toda},\ and\ \citenamefont {Hashitsume}}]{kubo}%
  \BibitemOpen
  \bibfield  {author} {\bibinfo {author} {\bibfnamefont {R.}~\bibnamefont
  {Kubo}}, \bibinfo {author} {\bibfnamefont {M.}~\bibnamefont {Toda}}, \ and\
  \bibinfo {author} {\bibfnamefont {N.}~\bibnamefont {Hashitsume}},\
  }\href@noop {} {\emph {\bibinfo {title} {Statistical Physics II}}},\ \bibinfo
  {edition} {2nd}\ ed.\ (\bibinfo  {publisher} {Springer},\ \bibinfo {year}
  {1998})\BibitemShut {NoStop}%
\bibitem [{\citenamefont {Reichl}(1998)}]{reichl}%
  \BibitemOpen
  \bibfield  {author} {\bibinfo {author} {\bibfnamefont {L.~E.}\ \bibnamefont
  {Reichl}},\ }\href@noop {} {\emph {\bibinfo {title} {A Modem Course in
  Statistical Physics}}},\ \bibinfo {edition} {2nd}\ ed.\ (\bibinfo
  {publisher} {Wiley},\ \bibinfo {year} {1998})\BibitemShut {NoStop}%
\bibitem [{\citenamefont {McQuarrie}(2000)}]{mcquarrie}%
  \BibitemOpen
  \bibfield  {author} {\bibinfo {author} {\bibfnamefont {D.~A.}\ \bibnamefont
  {McQuarrie}},\ }\href@noop {} {\emph {\bibinfo {title} {Statistical
  Mechanics}}}\ (\bibinfo  {publisher} {University Science Books},\ \bibinfo
  {address} {Sausalito},\ \bibinfo {year} {2000})\BibitemShut {NoStop}%
\bibitem [{\citenamefont {{Large-scale Atomic/Molecular Massively Parallel
  Simulator}}()}]{lammps}%
  \BibitemOpen
  \bibfield  {author} {\bibinfo {author} {\bibnamefont {{Large-scale
  Atomic/Molecular Massively Parallel Simulator}}},\ }\href@noop {} {}\bibinfo
  {note} {\url{http://lammps.sandia.gov}}\BibitemShut {NoStop}%
\bibitem [{\citenamefont {Plimpton}(1993)}]{plimpton1993fast}%
  \BibitemOpen
  \bibfield  {author} {\bibinfo {author} {\bibfnamefont {Steve}\ \bibnamefont
  {Plimpton}},\ }\href@noop {} {\emph {\bibinfo {title} {Fast parallel
  algorithms for short-range molecular dynamics}}},\ \bibinfo {type} {Tech.
  Rep.}\ (\bibinfo  {institution} {Sandia National Labs., Albuquerque, NM
  (United States)},\ \bibinfo {year} {1993})\BibitemShut {NoStop}%
\bibitem [{cca()}]{ccad}%
  \BibitemOpen
  \href@noop {} {}\bibinfo {note} {CCAD -- UNC:
  \url{http://ccad.unc.edu.ar/}}\BibitemShut {NoStop}%
\bibitem [{\citenamefont {{Maxima, a Computer Algebra System}}()}]{maxima}%
  \BibitemOpen
  \bibfield  {author} {\bibinfo {author} {\bibnamefont {{Maxima, a Computer
  Algebra System}}},\ }\href@noop {} {}\bibinfo {note}
  {\url{http://maxima.sourceforge.net/}}\BibitemShut {NoStop}%
\end{thebibliography}%

\end{document}